\documentclass[a4paper]{article}
\pdfoutput=1

\usepackage{physics}

\usepackage[utf8]{inputenc}
\usepackage[english]{babel}
\usepackage{tikz}
\usepackage{lipsum}

\usepackage{graphicx}  
\usepackage{amsmath}  
\usepackage{amssymb}  
\usepackage{hyperref}   
\usepackage{mathrsfs}   
\usepackage{gensymb}   
\usepackage{placeins}    
\usepackage{bm}           

\usepackage{authblk}    
\usepackage{array}       

\usepackage{pdflscape}   

\usepackage{tikz}           

\usepackage{algorithm}    
\usepackage{algorithmic}  

\usepackage{cite}

\hypersetup{
		breaklinks=true,  
		colorlinks = true,
		linkcolor={red}, 
		urlcolor={blue}, 
		citecolor={blue}
		}

\begin{document}

\title{
\vskip 2mm\bf\Large\hrule height5pt \vskip 4mm
Beyond QUBO and HOBO formulations, solving the Travelling Salesman Problem on a quantum boson sampler
\vskip 4mm \hrule height2pt
}

\author[1]{Daniel Goldsmith \thanks{daniel.goldsmith@digicatapult.org.uk}}
\affil[1]{Digital Catapult, 101 Euston Road, London NW1 2RA}
\author[1]{Joe Day-Evans}

\maketitle
\begin{tikzpicture}[remember picture,overlay]
   \node[anchor=north east,inner sep=0pt] at (current page.north east)
              {\includegraphics[scale=0.1]{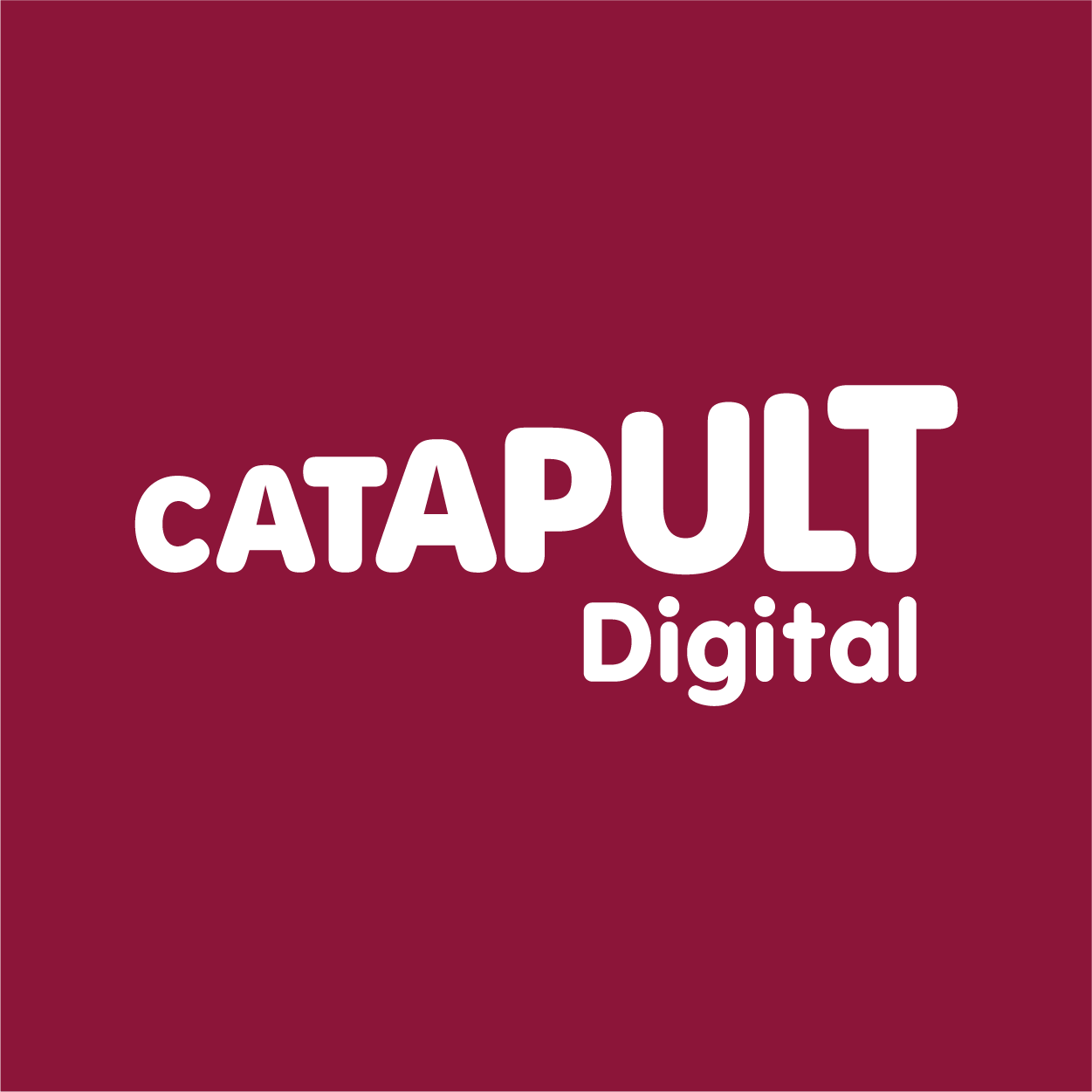}};
\end{tikzpicture}

\begin{abstract}
The Travelling Salesman Problem (TSP) is an important combinatorial optimisation problem, and is usually solved on a quantum computer  
using a Quadratic Unconstrained Binary Optimisation (QUBO) formulation 
or a Higher Order Binary Optimisation (HOBO) formulation.
In these formulations, penalty terms are added to the objective function for outputs that don't map to valid routes.

We present a novel formulation which needs fewer binary variables, and where, by design, 
there are no penalty terms because all outputs from the quantum device are mapped to valid routes.
Simulations of a quantum boson sampler were carried out which demonstrate that larger networks can be solved with this 
penalty-free formulation than with formulations with penalties.  
Simulations were successfully translated to hardware by running 
a non-QUBO formulation with penalties on an early experimental 
prototype ORCA PT-1 boson sampler.  Although we worked 
with a boson sampler, we believe that this novel formulation is relevant to other quantum devices.

This work shows that a good embedding for combinatorial optimisation 
problems can solve larger problems with the 
same quantum computing resource.    
The flexibility of boson sampling quantum devices is a powerful  
asset in solving combinatorial optimisation problem, 
because it enables formulations where the 
output string is always mapped to a 
valid solution, avoiding the need for penalties.

\end{abstract}

\section{Introduction}

We are in the `Noisy Intermediate Scale era of quantum computing' (NISQ), a phrase coined by Preskill \cite{Preskill1997}, 
and contemporary quantum devices have high error rates, limited connectivity and lack scale.  
Combinatorial optimization, where an objective function with binary arguments is minimised, has been identified as a potential early use case for 
quantum computing, along with quantum chemistry and quantum machine learning \cite{Bharti2021}.  

An important combinatorial optimisation problem is the well known NP-Hard Travelling Salesman Problem (TSP), which aims to find the shortest Hamiltonian cycle through a graph 
$G=(V, E)$ of $N$ locations represented as vertices $V$ in the graph, and edges $E$ between the vertices representing the connections.  
Each edge between vertices $u$ and $v$ has a distance $D_{u,v}$.
The TSP problem consists of finding a permutation $\pi$ of the $N$ locations, 
that minimises the distance $C(\pi)$ of visiting each of the $N$ locations once, and returning to the start.  The distance $C(\pi)$ is shown in Equation \ref{eq:1} where we 
note that $\pi(N+1) = \pi(1)$ because the salesman must return to the starting location.

\begin{equation} \label{eq:1} 
C(\pi) = \sum_{i=1}^{N}D_{\pi(i), \pi(i+1)} 
\end{equation}

As well as classical methods, quantum methods to solve TSP include quantum inspired algorithms \cite{Soloviev2021QuantumInspiredEO}, 
adiabatic quantum computing \cite{Jain2021SolvingTT, Kieu2018TheTS}, 
quantum heuristic based on a Grover search \cite{Bang2010AQH}, 
quantum phase estimation \cite{srinivasan2018efficient}, quantum approximate 
optimization algorithm (QAOA)\cite{ramezani2024reducing, Farhi2014, Ruan2020TheQA, Radzihovsky2019AQS} and 
using qudits rather than qubits \cite{Vargas_Calder_n_2021}.

The standard approach to solving this problem with a quantum computer is the Quadratic Unconstrained 
Binary Optimisation (QUBO) formulation as outlined by Lucas \cite{Lucas2013}, using either 
quantum annealing devices or the QAOA algorithm for circuit model devices.  In 2024 Ramezani and others \cite{ramezani2024reducing} designed and implemented a 
Higher Order Binary Optimisation (HOBO) algorithm for QAOA where the number of qubits scales 
as $\mathcal{O}(N\log_2{}N)$, rather than $\mathcal{O}(N^2)$ for the standard approach, but at the cost of increasing the 
number of 2-qubit gates and hence the circuit depth.  
In these QUBO and HOBO formulations, outputs that don't map to a valid solution attract penalty terms, 
and these invalid solutions may lead to a flat objective function that is difficult to optimise.  The QUBO and HOBO formulations are described in the Literature Review.

In the Methods section we present a novel formulation where, by design, 
there are no penalty terms because all outputs are mapped to valid routes, and explain how this formulation works with a boson sampler.  
This penalty-free formulation scales as $\mathcal{O}(N\log_2{}N)$ and requires slightly fewer binary variables than Ramezani's formulation \cite{ramezani2024reducing}, and 
we believe that it may be relevant for other quantum devices where the output is a binary string with a probability distribution 
depending on parameters that can be modified by a classical optimiser.

In the Experiments, Data and Results section we describe simulations of a quantum boson sampler that demonstrate that larger networks can be solved with this 
penalty-free formulation than with formulations with penalties, such as the QUBO formulation, or a formulation where the location 
reference is encoded in binary, similar to \cite{ramezani2024reducing}.  We also describe how we translated simulations to hardware by running 
a non-QUBO formulation with penalty terms on an early experimental 
prototype ORCA PT-1 boson sampler.   

\section{Literature review}

\subsection{QUBO formulation of TSP}

In the Quadratic Unconstrained Binary Optimisation (QUBO) formulation of a combinatorial optimisation problem 
the problem Hamiltonian $H_{P}$ is written as a function of binary variables
$x_i$ of no higher than quadratic order, as shown in Equation \ref{eq:combo}, and is minimised.  
The problem Hamiltonian contains  
the objective function to be minimised as well as penalty terms for any constraints, multiplied by arbitary constants to ensure that it is not 
energetically favourable to break the constraints:
\begin{equation} \label{eq:combo} 
H_{P} = \sum_{i,j} Q_{i,j} x_i x_j
\end{equation}
If the binary bit string is represented as a vector $\pmb{x}$ this is equivalent to the 
matrix formulation shown in Equation \ref{eq:matrix}:
\begin{equation} \label{eq:matrix} 
H_{P}= \pmb{x}^T \pmb{Q} \pmb{x}
\end{equation}
Lucas \cite{Lucas2013} describes a formulation of TSP as a QUBO.   
Labelling the vertices $0,1,...,N-1$, where $N$ is the number of vertices, or locations to be visited, 
$N^2$ binary variables $x_{v,i}$ are contemplated, where 
$v$ represents the vertex and $i$ its order in a prospective cycle.  
Since, without loss of generality, the cycle can always start at vertex $0$, the number of binary variables required is reduced to $\left( N - 1\right) ^ 2$.

As each edge between vertices $u$ and $v$ has a distance $D_{u,v}$ then the total distance is found 
as the sum of the distances for each edge: 
\begin{equation} \label{eq:9} 
H_{obj} =  \sum _{(uv) \in E}  D_{u,v} \sum_{j=0}^{N-1} x_{u,j}x_{v,j+1}
\end{equation}
Noting that the $v^{th}$ vertex can only appear once in the cycle, and 
that there must be a $j^{th}$ node in the cycle for each $j$, 
these constraints are encoded in the penalty Hamiltonian:
\begin{equation} \label{eq:8} 
H_{pen} = A\sum_{v=0}^{N-1}\left( 1 - \sum_{j=0}^{N-1} x_{v,j}\right)^2 + A\sum_{j=0}^{N-1}\left( 1 - \sum_{v=0}^{N-1} x_{v,j}\right)^2
\end{equation}
with $A$ large enough so that it is never favourable to violate the constraints of $H_{pen}$.  

In summary, in the QUBO formulation of TSP, we minimise the problem Hamiltonian:
\begin{equation} \label{eq:10} 
H_{P} =  H_{obj} + H_{pen}
\end{equation}

\subsubsection{Solving the QUBO formulation of TSP on a quantum annealing device}
The QUBO formulation in \ref{eq:10} can be solved with a quantum annealing device using adiabatic quantum computing 
because the quadratic problem Hamiltonian $H_{P}$ in Equation \ref{eq:10} can be mapped to the Hamiltonian for a quantum annealing device as shown in Equation \ref{eq:6} which is also quadratic:

\begin{equation} \label{eq:6} 
H_{P} = \sum_i h_i \sigma_{z}^{(i)} + \sum_{i,j} J_{i,j} \sigma_{z}^{(i)} \sigma_{z}^{(j)}
\end{equation}
where $\sigma_{z}^{(i)}$ denotes the Pauli-Z matrix acting on the $i^{th}$ qubit, the $h_i$ biases are individually set for each qubit, 
and the $J_{i,j}$ biases control the strength of  interactions between the $i^{th}$ and $j^{th}$ qubit.  
Since each binary variable in the problem is mapped to one qubit, and more qubits 
may be needed if one qubit can't be connected to another, the problem requires at least  $\mathcal{O}(N^2)$ qubits

Only small TSP networks can be solved at present using Quantum Annealing.  For example, in 2021 Jain \cite{Jain2021SolvingTT} 
struggled to solve networks of more than 10 locations with the D-Wave quantum annealer.

\subsubsection{Solving the QUBO formulation of TSP on a circuit model device}
In the QAOA algorithm \cite{Farhi2014} the time evolution of the adiabatic quantum computing Hamiltonian 
is Trotterised using a paramaterized quantum circuit which encodes a Unitary operator
 $U(\bm{\gamma}, \bm{\beta})$, where $\bm{\gamma}$ and $\bm{\beta}$ are sets of variational parameters.

The Unitary operator takes the form shown in Equation \ref {eq:qaoa1}
\begin{equation} \label{eq:qaoa1} 
	U(\bm{\gamma}, \bm{\beta}) = \prod_{l=1}^{L} \exp\left(-i\beta_{l}H_{M}\right)\exp\left(-i\gamma_{l}H_{P}\right)
\end{equation}
where $L$ denotes the number of optimization layers. The problem Hamiltonian $H_P$ is as described in Equation \ref {eq:6} 
and the $H_M$ is a mixing Hamiltonian.  

\subsection{HOBO formulation of TSP}
The QUBO formulation above uses a One-Hot encoding scheme.  Only a few combinations of binary 
variables are valid solutions because if the binary variables $x_{v,i}$ are 
represented as a matrix, few matrices are valid permutation matrices. It is reasonable to ask if we can do better with 
a Higher Order Binary Optimisation (HOBO) formulation.
 
In 2021 Vargas-Calderon and others \cite{Vargas_Calder_n_2021} formulated TSP using qudits rather than qubits.  
They encoded a valid route,  for example, $0\to3\to2\to1\to0$, using an ordered set of four 4-level qudits
 as a quantum state $\ket{0}\otimes\ket{3}\otimes\ket{2}\otimes\ket{1}$.  
They devised a Hamiltonian that returns the total distance for valid cycles, and a penalty term otherwise, 
and is a sum of quadratic terms involving no more than two qudits.
The team simulated valid solutions for up to 40 locations, although they could not yet run this formulation on an existing device, 
since contemporary devices use qubits, not qudits.

In 2024 Ramezani and others \cite{ramezani2024reducing} built on Vargas-Calderon's work by introduced a HOBO formulation.  
Routes are encoded using the binary representation of 
the location number.  For example, a valid cycle $0\to3\to2\to1\to0$ is 
encoded as a binary string of ${00 11 10 01}_2$ = ${00}_2$ ${11}_2$ ${10}_2$ ${01}_2$.  

Since each binary variable is mapped to a qubit, this reduces the qubits required to $N\left\lceil \log_2{}N \right\rceil$, 
since there are $N$ locations, and each location reference requires $\left\lceil \log_2{}N \right\rceil$ binary variables. 
In the problem Hamiltonian the reduction in the number of binary variables comes at the expense of introducing higher order binary terms. 
Because of this the number of two-qubit gates scales as  $\mathcal{O}(N^4\log_2{}N)$ rather than $\mathcal{O}(N^3)$ for 
the conventional QAOA algorithm.  
Simulation by Ramezani's team showed that the HOBO formulation returned better solutions than standard methods, but only 
for small networks with four locations.  
 
\section{Methods}

\subsection{Formulation of the TSP without penalty terms}

A novel formulation is presented, which unlike the QUBO and HOBO formulations above, does not use penalty terms.  
Although this formulation was implemented on a quantum boson sampler, we believe the penalty free
 formulation may be relevant for other quantum devices where the output is a binary string with a probability distribution 
depending on parameters that can be modified by a classical optimiser.

An initial sequential list of location numbers is created and items are moved from the initial list to a re-ordered list until the initial list is empty and 
the re-ordered list holds a valid cycle.  To achieve this the output bit string from the quantum device is split into smaller strings, 
which are interpreted as the binary encoding of indices pointing to the next item of the ordered list to be moved.    
Use of modulo arithmetic 
prevents errors where the index is too high to reference a valid element of the ordered list.  

The required bit string length $L_1$ is found by summing the length of these binary index strings, 
noting that the length of the string is $\left\lceil \log_2(i) \right\rceil$  where $i$ is 
the number of elements remaining in the ordered list.  
$L_1$ is given by Equation \ref{eq:len2} where $N$ is the number of locations:
\begin{equation} \label{eq:len2} 
L_1 = \sum _{i=1} ^{N-1} \left\lceil \log_2(i) \right\rceil 
\end{equation}
The number of binary variables $L_1$ required is less than the $N\left\lceil \log_2{}N \right\rceil$ 
 for Ramezani's formulation \cite{ramezani2024reducing}, because this encoding is more efficient than Ramezani's
binary conversion of each location code, and substantially 
less than the $\left( N - 1\right) ^ 2$ needed for the standard QUBO formulation \cite{Lucas2013}. 

Pseudo code for implementation of this algorithm on a boson sampler is shown as Algorithm \ref {algo:1} in the \hyperlink{Appendix}{Appendix}.  
This formulation was simulated on a boson sampling device, as described below:  

\subsection{Optimisation on a boson sampling device}
In a photonic boson sampler indistinguishable photons are injected into a 
linear interferometer and the output distribution is measured.  
The output photon probability distribution can be shown to be hard to simulate classically \cite{Aaronson2010TheCC}, a necessary condition for quantum advantage.
  Bradler and  Wallner \cite{Bradler2021} explain 
how a photonic boson sampling device can carry out optimisation using Variational Quantum Algorithms (VQAs) \cite{Bharti2021},
and illustrate this by solving a 
Mobius graph problem in a QUBO formulation, 
and a portfolio optimization problem with binary investment problem in a non-QUBO formulation.

\begin{figure}[h!]
    \centering
    \includegraphics[width=1\textwidth]{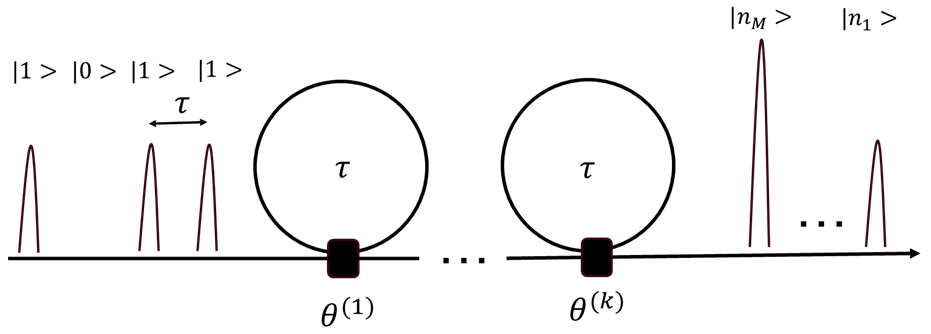}
    \caption{The ORCA PT-Series following \cite{Bradler2021} with a train of optical pulses that interfere with each other via programmable beam splitters
parameterised by the beam splitter angles $\theta^{(i)}$}
    \label{fig:splitter}
\end{figure}

An example of a photonic boson sampler is the ORCA Computing PT-Series quantum computer, which produces of a train of optical pulses that interfere with each other 
via programmable beam splitters located within one or more fibre loops, with a parameter vector given by the beam splitter angles $\pmb\theta = \theta^{(i)}$,
as in Figure \ref{fig:splitter}.

\begin{figure}[h!]
    \centering
    \includegraphics[width=1\textwidth]{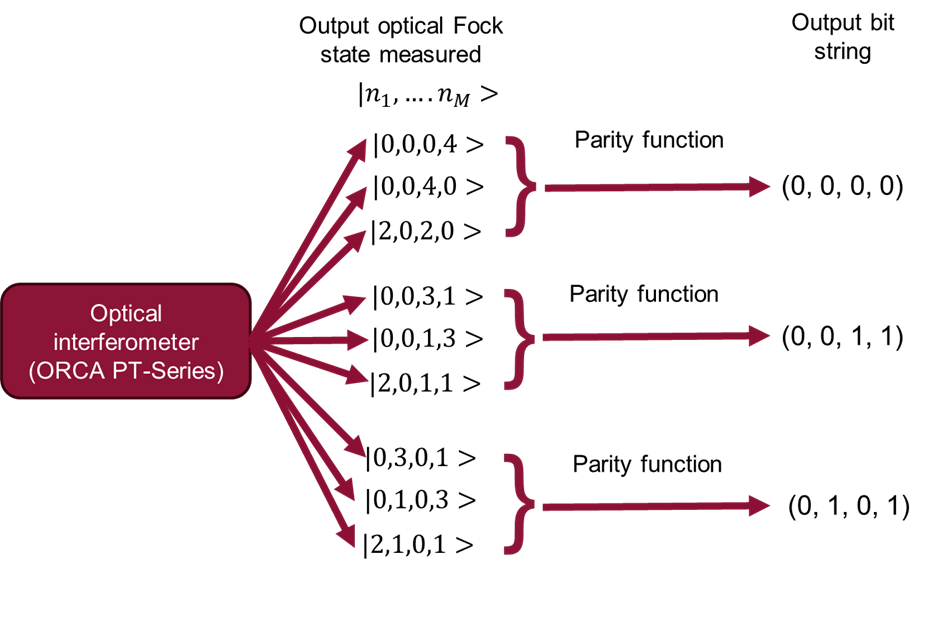}
    \caption{Component wise parity function from optical circuit outputs to bit strings based on \cite{Bradler2021}. }
    \label{fig:parity}
\end{figure}

The output of this boson sampler, a quantum Fock state, with an integer number of photons in each output position, 
is measured and then mapped to a bit string using component wise parity functions.  
An example of a component-wise parity function mapping from the optical circuit outputs to bit strings  
 is show in Figure \ref{fig:parity} with 
even or odd photon counts mapped to $0$ or $1$ \cite{Bradler2021}.  
It can shown that the device needs to be run in four configurations, corresponding to sampling steps for odd and even parity counts and for $M$ and $M-1$ photons, so that all bit strings can be sampled.

The objective function is evaluated classically for each output bit string and then optimised classically 
by tuning the parameter vector of beam splitter angles $\pmb\theta = \theta^{(i)}$ in a feedback loop, which is repeated 
until the objective function converges, at which point the lowest energy found 
and the corresponding bit string found are recorded.

\FloatBarrier

\subsection{Comparison with other formulations}
The performance of the penalty-term free formulation on the bosom sampler was assessed by comparing it to a non-QUBO and a QUBO formulation, both with penalty terms

\subsubsection{Non-QUBO formulation of the TSP with penalty terms on a boson sampler}

Using the flexibility of the boson sampler, Ramezani's \cite{ramezani2024reducing} technique can be used to encode 
the boson sampler output binary string as a location list, using a binary representation of the location label, but without explicitly defining the full HOBO formulation.  
If the location list contains  
duplicate location labels, or location labels higher than $(N-1)$ then a constant penalty term $H_{pen}$, 
as defined in Equation \ref {eq:pen1}, is applied with a value of $\rho$ high enough 
so that these terms are not selected in preference to a valid cycle.  In the simulations a value of $\rho = 5$ was used.  
Otherwise the location list defines a valid cycle, and the cycle distance is calculated.
\begin{equation} \label{eq:pen1} 
H_{pen}  = \rho \sum_{i=0} ^{N-1} D_{0,i}
\end{equation}

Pseudo code is shown as Algorithm \ref{algo:2} in the \hyperlink{Appendix}{Appendix}.
This formulation was simulated on a boson sampling device and also run on the ORCA PT-1 device.

Whilst this formulation was able to find solutions for up to 15 locations, as discussed in the \hyperlink{results}{results}
below, the solutions were not high quality.  The reason seems likely to be that many of the bit strings 
result in the penalty function being applied, so the cost landscape is flat, and the optimiser struggles to find the minimum.  

\subsubsection{QUBO formulation of the TSP with penalty terms on a boson sampler}

A boson sampler output binary string of length $L_3 = (N-1)^2$ was reshaped as a matrix with elements $x_{v,j}$, to reproduce the QUBO formulation.  Distances were calculated for 
binary strings representing valid routes, and penalty terms were applied if the route was not valid.
Pseudo code is shown as Algorithm \ref{algo:3} in the \hyperlink{Appendix}{Appendix}.
This formulation was simulated on a boson sampling device. More binary variables were required, and only small networks could be solved, as discussed in the \hyperlink{results}{results} below.

\section{Experiments, data and results}

\subsection{Experiments carried out}

The three boson sampler TSP formulations described above were simulated on a boson sampler using the ORCA SDK with networks of between four 
and 48 locations.  The solution quality ($Q_{sol}$) was determined as 
the shortest distance found experimentally around a valid cycle ($D_{min\_exp}$) as a percentage of the best known minimum distance ($D_{min\_best}$) as shown in Equation \ref{eq:quality}

\begin{equation} \label{eq:quality} 
Q_{sol}  =  100\frac{D_{min\_exp}}{D_{min\_best}}
\end{equation}

A solution quality of $Q_{sol} = 100\%$ represents a solution where the shortest distance found is optimal, and a lower percentage represents a sub-optimal solution.

Key training graphs are presented showing how the objective function varies with the number of training iterations.  
Four different plots are shown on each graph represent the four sampling configurations for odd and even parity counts and for $M$ and $M-1$ photons, as explained in the 
description of the boson sampling device. 

The simulations were run on a laptop with an Intel(R) Core(TM) i5-10210U CPU clocked at 1.60GHz with 8.00 GB (7.76 GB usable) RAM running Windows 11. 
We used the latest ORCA SDK which can be obtained from ORCA Computing.  A link to an earlier version of the SDK can be found in \cite{Bradler2021}.  
We have placed the relevant Jupyter Notebooks and Python modules on a Digital Catapult GitHub repository \cite{DC_Github}. 

To assess the impact of hyper-parameters, simulations were run with 
different numbers of iterations and learning rates.  
The optimisation method was either the parameter shift rule, 
or Simultaneous Perturbation Stochastic Approximation (SPSA) \cite{SPSA}, where instead of estimating the 
gradient for each parameter by taking two function estimates per parameter, 
the gradient was estimated with two function estimation steps based on a stochastic variation.  

The non-QUBO based formulation with penalty terms was also run on an experimental ORCA PT-1 boson sampler.  
Since this early device could only accommodate six binary variables it was run for four locations.    

\subsection{Data}
Data sets with $5$, $15$, $26$, $42$ and $48$ locations were obtained 
from \cite{tspdata, Rein91}, and a data set with 
$4$ locations was used for Unit Testing, where the results of the simulations were checked manually. 

\hypertarget{results}{\subsection{Results}}

The non-QUBO formulation with penalties was run on a network of 4 locations on an experimental ORCA-PT1 boson sampler, and the correct solution was found.

Results of simulation with different formulations, numbers of locations, iterations, learning rates and optimisation methods are summarised in the table below:

\begin{table}[h!]
\centering
\begin{tabular}{c}
    \includegraphics[width=1\textwidth]{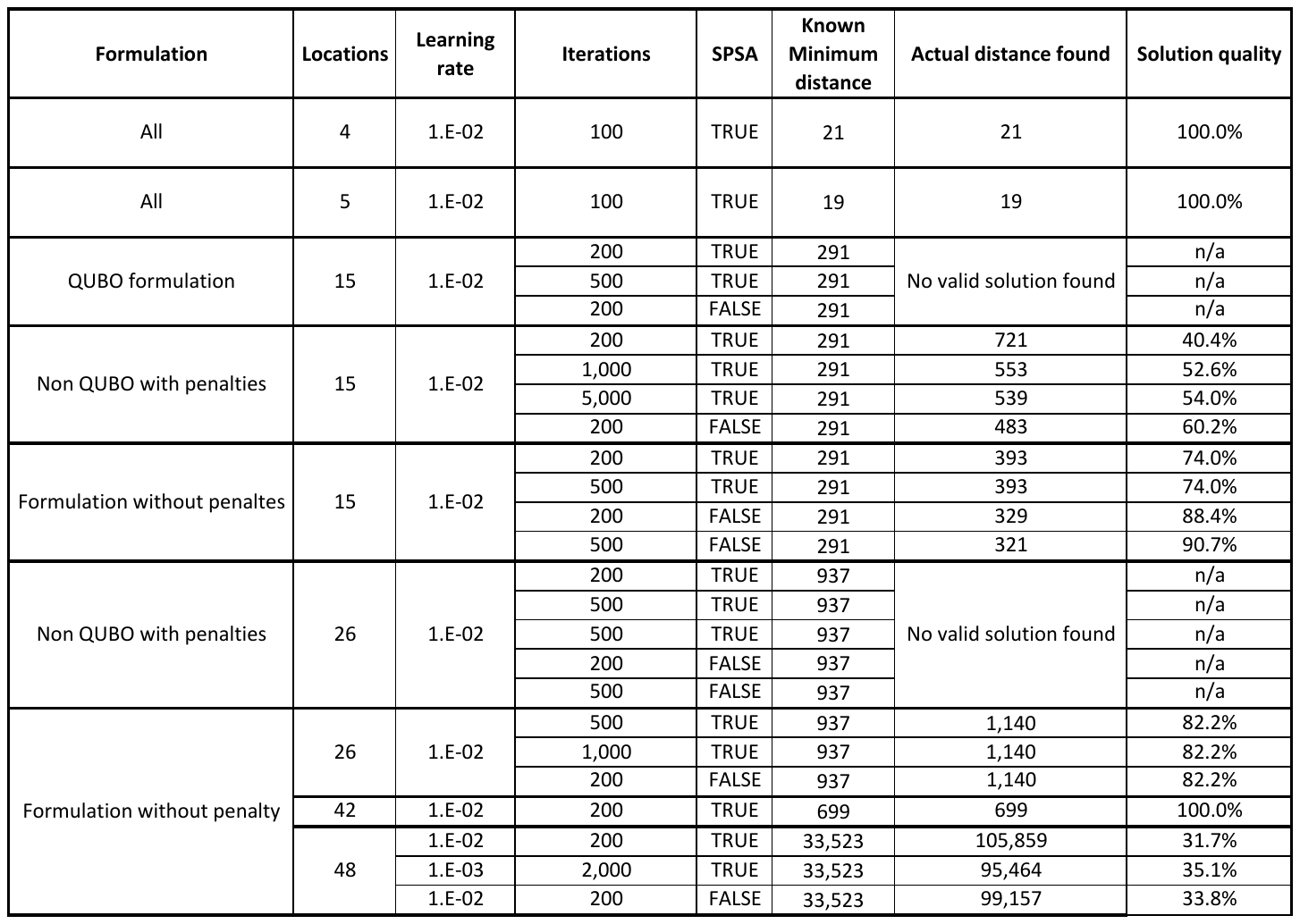}
\end{tabular}
\caption{Summary of simulation results}
\label{table:results}
\end{table}

\FloatBarrier

\subsubsection{Simulation of small networks}

Each of the three formulations simulated found the correct solution for networks of $4$ or $5$ locations.
 
\subsubsection{Simulations of a network with 15 locations}
The QUBO formulation could not find a valid solution for 15 locations.  We believe this is 
because almost all binary strings were outside the valid solution space, and attracted the same penalty 
term shown in Equation \ref{eq:pen1}, leading to a flat cost landscape, where it is difficult for the optimiser to find a minimum.    

\begin{figure}[h!]
    \centering
    \includegraphics[width=0.8\textwidth]{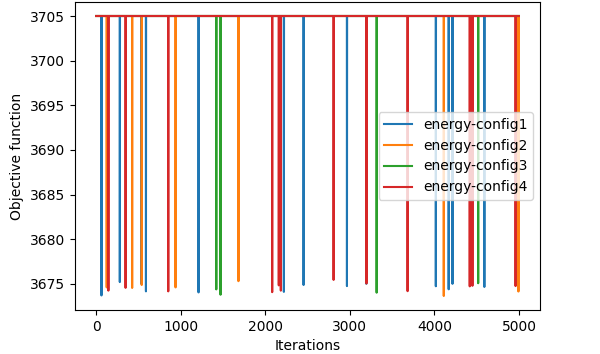}
    \caption{Objective function against iterations for non-QUBO formulation with penalties showing that because almost 
all bit strings attract the same penalty term, no convergence is possible on a network with 15 locations.  The four plots represent the four sampling configurations.}
    \label{fig:results1}
\end{figure}

Both non-QUBO formulations  
found valid solution for 15 locations.  However, the training graph with SPSA and a learning rate of 0.01
for the non-QUBO formulation with penalties in Figure \ref{fig:results1} 
shows that most bit strings have a cost equal to the penalty term, apart from a few strings with a valid solution.  
The average objective function does not reduce with the number of iterations, and there is no convergence to an optimum solution.  

The penalty free formulation solution quality was 
higher for 15 locations that the formulation with penalties, with a solution quality of up to $90.7\%$.  

\subsubsection{Simulations of network with over 26 locations}

\begin{figure}[h!]
    \centering
    \includegraphics[width=0.8\textwidth]{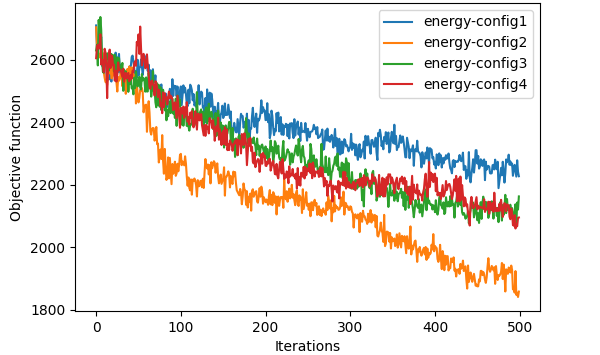}
    \caption{Objective function against iterations for the penalty-free formulation showing evidence of training for a network of 26 locations.  The four plots represent the four sampling configurations.}
    \label{fig:results2}
\end{figure}

The non-QUBO formulation with penalties could not find a valid solution for 26 locations.  
The penalty-free formulation found solutions with up to 48 locations, even though 
the solution quality declined with larger networks.    
The penalty-free formulation showed evidence of training with the value of the objective function decreasing during training.
This is clearly seen with 26 locations, SPSA and a learning rate of 0.01, 
as shown in Figure \ref{fig:results2}, although it was less pronounced in other network sizes.   

\subsubsection{Impact of optimisation method}

SPSA was generally much quicker, presumably because only two estimates were made, 
and was noiser, usually with a lower solution quality.

\FloatBarrier

\section{Discussion}

Simulations of a novel, penalty free formulation of TSP on a boson sampler has shown useful results, 
with valid solutions for networks of up to 48 locations, even though 
the solution quality declined with larger networks.  The use of penalties may be associated with flat, difficult to optimise, 
training graphs.

Whilst the problem size and solution quality is not yet comparable to best of breed classical methods, 
the simulation carried out was of an early boson 
sampler with only one fibre loop to store photons, and more advanced devices are likely soon.    
Also, much information from the measurement of the Fock space is discarded, 
since integer values are mapped to binary parities.  
Accessing these integers may lead to an improved formulation.

Future work is needed to investigate the performance of this penalty free formulation 
by simulating and/or running it on larger and more complex boson samplers, or other devices.

\section{Conclusions}

This work shows that: 
\begin{enumerate}
   \item It is important to use a good embedding for TSP, 
and by inference, other combinatorial problems.  
   \item An embedding designed to 
always map the binary output string of a quantum device to a  
valid solution, avoiding the need for penalty terms, was able to solve larger routes than formulations with penalties.   

\end{enumerate}

\section{Acknowledgments}
We are very grateful to Dr Clive Emary from Frazer-Nash and Dr William Clements from ORCA for their invaluable contribution to this work.  
Many thanks to the ORCA Computing 
team for allowing access to an ORCA PT-1 experimental prototype to test the algorithm on a real device, and allowing access to the 
latest version of their software development kit (SDK).
Many thanks to Sheevan Jayasinghe from Bahut, a QTAP participant, 
who contributed the TSP use case, and an example coded in the QUBO formulation.

This work was carried out during the use case dissemination phase of 
the Quantum Technology Access Programme (QTAP), 
part of the Quantum Data Centre of the Future, project number 10004793,  
funded by the Innovate UK Industrial Strategy Challenge Fund (ISCF).  

\bibliographystyle{unsrt} 
\bibliography{tsp}

\begin{thebibliography}{10}

\bibitem{Preskill1997}
John Preskill.
\newblock Fault-tolerant quantum computation.
\newblock
  \href{https://arxiv.org/abs/quant-ph/9712048}{https://arxiv.org/abs/quant-ph/9712048},
  12 1997.

\bibitem{Bharti2021}
Kishor Bharti, Alba Cervera-Lierta, Thi~Ha Kyaw, Tobias Haug, Sumner
  Alperin-Lea, Abhinav Anand, Matthias Degroote, Hermanni Heimonen, Jakob~S.
  Kottmann, Tim Menke, Wai-Keong Mok, Sukin Sim, Leong-Chuan Kwek, and Alán
  Aspuru-Guzik.
\newblock Noisy intermediate-scale quantum (nisq) algorithms.
\newblock {\em Reviews of Modern Physics}, 94, 1 2021.

\bibitem{Soloviev2021QuantumInspiredEO}
Vicente~P. Soloviev, Concha Bielza, and Pedro Larra{\~n}aga.
\newblock Quantum-inspired estimation of distribution algorithm to solve the
  travelling salesman problem.
\newblock {\em 2021 IEEE Congress on Evolutionary Computation (CEC)}, pages
  416--425, 2021.

\bibitem{Jain2021SolvingTT}
Siddhartha Jain.
\newblock Solving the traveling salesman problem on the d-wave quantum
  computer.
\newblock In {\em Frontiers of Physics}, 2021.

\bibitem{Kieu2018TheTS}
Tien~D. Kieu.
\newblock The travelling salesman problem and adiabatic quantum computation: an
  algorithm.
\newblock {\em Quantum Information Processing}, 18:1--19, 2018.

\bibitem{Bang2010AQH}
Jeongho Bang, Junghee Ryu, Changhyoup Lee, Seokwon Yoo, James Lim, and
  Jinhyoung Lee.
\newblock A quantum heuristic algorithm for the traveling salesman problem.
\newblock {\em Journal of the Korean Physical Society}, 61:1944 -- 1949, 2010.

\bibitem{srinivasan2018efficient}
Karthik Srinivasan, Saipriya Satyajit, Bikash~K. Behera, and Prasanta~K.
  Panigrahi.
\newblock Efficient quantum algorithm for solving travelling salesman problem:
  An ibm quantum experience.
\newblock
  \href{https://arxiv.org/abs/1805.10928}{https://arxiv.org/abs/1805.10928},
  2018.

\bibitem{ramezani2024reducing}
Mehdi Ramezani, Sadegh Salami, Mehdi Shokhmkar, Morteza Moradi, and Alireza
  Bahrampour.
\newblock Reducing the number of qubits from $n^2$ to $n\log_{2} (n)$ to solve
  the traveling salesman problem with quantum computers: A proposal for
  demonstrating quantum supremacy in the nisq era.
\newblock
  \href{https://arxiv.org/abs/2402.18530}{https://arxiv.org/abs/2402.18530},
  2024.

\bibitem{Farhi2014}
Edward Farhi, Jeffrey Goldstone, and Sam Gutmann.
\newblock A quantum approximate optimization algorithm.
\newblock
  \href{http://arxiv.org/abs/1411.4028}{http://arxiv.org/abs/1411.4028}, 11
  2014.

\bibitem{Ruan2020TheQA}
Yue Ruan, Samuel Marsh, Xiling Xue, Zhihao Liu, and Jingbo~B. Wang.
\newblock The quantum approximate algorithm for solving traveling salesman
  problem.
\newblock {\em Computers, Materials \& Continua}, 2020.

\bibitem{Radzihovsky2019AQS}
Matthew Radzihovsky, Joey Murphy, and Mason Swofford.
\newblock A qaoa solution to the traveling salesman problem using pyquil.
\newblock
  \href{https://cs269q.stanford.edu/projects2019/radzihovsky_murphy_swofford_Y.pdf}{https://cs269q.stanford.edu/projects2019/radzihovsky\_murphy\_swofford\_Y.pdf},
  2019.

\bibitem{Vargas_Calder_n_2021}
Vladimir Vargas-Calderón, Nicolas Parra-A., Herbert Vinck-Posada, and Fabio~A.
  González.
\newblock Many-qudit representation for the travelling salesman problem
  optimisation.
\newblock {\em Journal of the Physical Society of Japan}, 90(11):114002,
  November 2021.

\bibitem{Lucas2013}
Andrew Lucas.
\newblock Ising formulations of many np problems.
\newblock
  \href{https://arxiv.org/abs/1302.5843}{https://arxiv.org/abs/1302.5843}, 2
  2013.

\bibitem{Aaronson2010TheCC}
Scott Aaronson and Alexei~Y. Arkhipov.
\newblock The computational complexity of linear optics.
\newblock
  \href{https://arxiv.org/abs/1011.3245}{https://arxiv.org/abs/1011.3245},
  2010.

\bibitem{Bradler2021}
Kamil Bradler and Hugo Wallner.
\newblock Certain properties and applications of shallow bosonic circuits.
\newblock
  \href{https://arxiv.org/abs/2112.09766}{https://arxiv.org/abs/2112.09766}, 12
  2021.

\bibitem{DC_Github}
Daniel Goldsmith.
\newblock Quantum technology access programme software repository.
\newblock
  \href{https://github.com/digicatapult/QTAP}{https://github.com/digicatapult/QTAP},
  2023.

\bibitem{SPSA}
James~C. Spall.
\newblock Simultaneous perturbation stochastic approximation: Spsa - a method
  for system optimisation.
\newblock \href{https://www.jhuapl.edu/SPSA/}{https://www.jhuapl.edu/SPSA/},
  2001.

\bibitem{tspdata}
Tsp - data for the traveling salesperson problem.
\newblock
  \href{https://people.sc.fsu.edu/~jburkardt/datasets/tsp/tsp.html}{https://people.sc.fsu.edu/~jburkardt/datasets/tsp/tsp.html},
  2019.
\newblock Reviewed 13 February 2024.

\bibitem{Rein91}
Gerhard Reinelt.
\newblock {TSPLIB}--a traveling salesman problem library.
\newblock {\em ORSA Journal on Computing}, 3(4):376--384, 1991.

\end{thebibliography}

\hypertarget{Appendix}{\section*{Appendix}} 

\subsection*{Algorithms}

In these algorithms we follow the convention from Python that the first item of a list is indexed by $0$.

\begin{algorithm}
	\caption{Formulation of TSP without penalty terms on a boson sampler}
	\label{algo:1}
	\begin{algorithmic}
	\STATE \textbf{Input}: 
	\STATE Number of locations $N$
         \STATE Distance matrix $D_{u,v}$
         \STATE Number of samples $N_s$
	\STATE
	\STATE \textbf{Output}: 
	\STATE Minimum distance $D_{min}$
	\STATE Cycle with minimum distance $LIST_{min}$
	\STATE
	\STATE Calculate the bit string length $L_1$ from Equation \ref{eq:len2}
	\STATE Configure the interferometer to produce bit strings $\pmb b$ of length $L_1$
	\FOR{$i \gets 1$ to $N_s$}
	\STATE Measure the output pattern generated by the interferometer.
	\STATE Classical map the output pattern to a bit string $\pmb b$ of length $L_1$
	\STATE Initialise an ordered list of locations $LIST_{ordered} = [0, 1, ... N-1]$ 
	\STATE Initialise an empty list of locations $LIST_{cycle} = [\:]$ 
	\STATE Move the $0^{th}$ element $0$ from $LIST_{ordered}$ to $LIST_{cycle}$ since, without loss of generality, the cycle can start at location $0$.
	\FOR{$j \gets 1$ to $N-2$}
	\STATE Calculate the number of items $m_j$ left in $LIST_{ordered}$ as $m_j= N - j$
	\STATE Calculate the length $l_j$ of bit string $\pmb b_j$ as  $l_j = \left\lceil log_2(m_j) \right\rceil$
	\STATE Move $l_j$ bits from $\pmb b$ to $\pmb b_j$
	\STATE Evaluate $\pmb b_j$ as an integer $p_j$
	\STATE Calculate an index $q_j = p_j mod(m_j)$  
	\STATE Move the $q_j^{th}$ element from $LIST_{ordered}$ to $LIST_{cycle}$
	\ENDFOR
	\STATE Move the final element $LIST_{ordered}$ to $LIST_{cycle}$
	\STATE Calculate the distance $D_i$ for cycle $LIST_{cycle}$ using the distance matrix $D_{u,v}$
	\IF {$D_{min}$ is initial} 
	\STATE $D_{min} \gets D_i$ 
	\ELSIF{$D_i < D_{min}$} 
	\STATE $D_{min} \gets D_i$ 
	\STATE $LIST_{min} \gets LIST_{cycle} $
	\ENDIF
	\ENDFOR
	\STATE Iterate changing the parameter vector $\pmb\theta$ until convergence is reached.
	\end{algorithmic}
\end{algorithm}

\begin{algorithm}
	\caption{Non-QUBO formulation of the TSP with penalty terms on a boson sampler}
	\label{algo:2}
	\begin{algorithmic}
	\STATE \textbf{Input}: 
	\STATE Number of locations $N$
         \STATE Distance matrix $D_{u,v}$
         \STATE Number of samples $N_s$
	\STATE
	\STATE \textbf{Output}: 
	\STATE Bit string for which a minimum distance is found $ \pmb b_{min}$
	\STATE Minimum problem Hamiltonian $H_{min}$
	\STATE Minimum distance $D_{min}$
	\STATE Cycle with minimum distance $LIST_{min}$ 
	\STATE
	\STATE Calculate the bit string length $L_2 = \left\lceil log_2(N) \right\rceil (N-1)$
	\STATE Configure the interferometer to produce bit strings $\pmb b$ of length $L_2$
	\FOR{$i \gets 1$ to $N_s$}
	\STATE Measure the output pattern generated by the interferometer.
	\STATE Initialise an empty list of locations $LIST_{cycle} = [\:]$ 
	\FOR{$j \gets 1$ to $N-2$}
	\STATE Move $\left\lceil log_2(N) \right\rceil$ bits from $\pmb b$ to $\pmb b_j$
	\STATE Evaluate $\pmb b_j$ as an integer $p_i$
	\STATE Append $p_i$ to $LIST_{cycle}$ 
	\ENDFOR
	\STATE Augment $LIST_{cycle}$ by adding the missing location by elimination
	\IF{$LIST_{cycle}$ is valid}
	\STATE Calculate the distance $D_i$ for cycle $LIST_{cycle}$ using the distance matrix $D_{u,v}$
	\STATE $H_{problem} \gets D_i$
	\ELSE
	\STATE $H_{problem} \gets H_{pen}$
	\ENDIF
	\IF {$H_{min}$ is initial} 
	\STATE $H_{min} \gets H_{problem}$
	\ELSIF{$H_{problem} < H_{min}$} 
	\STATE $H_{min} \gets H_{problem}$
	\STATE $\pmb b_{min} \gets \pmb b$
	\ENDIF
	\ENDFOR
	\STATE Iterate, changing the parameter vector $\pmb\theta$ until convergence is reached.
	\STATE Calculate $LIST_{min}$ from $\pmb b_{min}$
	\STATE Calculate $D_{min}$ from $LIST_{min}$ and $D_{u,v}$
	\end{algorithmic}
\end{algorithm}

\begin{algorithm}
	\caption{QUBO formulation of TSP on a boson sampler.}
	\label{algo:3}
	\begin{algorithmic}
	\STATE \textbf{Input}: 
	\STATE Number of locations $N$
         \STATE Distance matrix $D_{u,v}$
         \STATE Number of samples $N_s$
	\STATE
	\STATE \textbf{Output}: 
	\STATE Bit string for which a minimum distance is found $\pmb b_{min}$
	\STATE Minimum problem Hamiltonian $H_{min}$
	\STATE Minimum distance $D_{min}$
	\STATE Cycle with minimum distance $LIST_{min}$ 
	\STATE
	\STATE Calculate the bit string length $L_3$ as $(N-1)^2$
	\STATE Configure the interferometer to produce bit strings $\pmb b$ of length $L_3$
	\FOR{$i \gets 1$ to $N_s$}
	\STATE Measure the output pattern generated by the interferometer.
	\STATE Classical map the output pattern to a bit string $\pmb b$ of length $L_3$
	\STATE Reshape $\pmb b$ to a matrix with elements $x_{i,j}$
	\STATE Calculate $H_{problem}$ classically using Equation \ref{eq:10} and $D_{u,v}$
	\IF {$H_{min}$ is initial} 
	\STATE $H_{min} \gets H_{problem}$
	\ELSIF{$H_{problem}< H_{min}$} 
	\STATE $H_{min} \gets H_{problem}$
	\STATE $\pmb b_{min} \gets \pmb b$
	\ENDIF
	\ENDFOR
	\STATE Iterate changing the parameter vector $\pmb\theta$ until convergence is reached.
	\STATE Calculate $LIST_{min}$ from $\pmb b_{min}$
	\STATE Calculate $D_{min}$ from $LIST_{min}$ and $D_{u,v}$
	\end{algorithmic}
\end{algorithm}

\end{document}